\documentclass[a4paper,UKenglish]{oasics}

\usepackage{microtype}

\usepackage[all]{xy}
\usepackage{wrapfig}

\newcommand{\TT}{\ensuremath{{\mathcal T}^\DB_\eta}}
\newcommand{\T}{\ensuremath{\mathcal T}}
\newcommand{\DB}{\ensuremath{\mathit{DB}}}
\newcommand{\OO}{\ensuremath{\mathcal O}}
\newcommand{\U}{\ensuremath{\mathcal U}}
\newcommand{\V}{\ensuremath{\mathcal V}}
\newcommand{\body}{\ensuremath{\mathsf{body}}}
\newcommand{\head}{\ensuremath{\mathsf{head}}}
\newcommand{\lit}{\ensuremath{\mathsf{lit}}}
\newcommand{\nup}{\ensuremath{\mathsf{nup}}}
\newcommand{\At}{\ensuremath{\mathit{At}}}
\newcommand{\normalize}{\ensuremath{\mathcal N}}
\newcommand{\Ieta}[1][]{\langle\DB,\eta_{#1}\rangle}
\newcommand{\neff}{\ensuremath{\mathsf{neff}_{\DB}}}
\newcommand{\ua}{\ensuremath{\mathsf{ua}}}
\newcommand{\indep}{\mathrel{\raisebox{-1mm}{\rotatebox{90}{$\models$}}}}
\newcommand{\ii}{\ensuremath{\hat\imath}}
\newcommand{\nondet}[1]{\ensuremath{#1^{\uparrow}}}
\newcommand{\neta}{{\normalize(\eta)}}

\title{Grounded Fixpoints and Active Integrity Constraints\footnote{Supported by the Danish Council for Independent Research, Natural Sciences, grant DFF-1323-00247.}}

\author{Lu\'{\i}s Cruz-Filipe}

\affil{Dept.\ Mathematics and Computer Science, University of Southern Denmark\\
  Campusvej 55, 5230 ODENSE M, Denmark\\
  \texttt{lcfilipe@gmail.com}}

\authorrunning{L. Cruz-Filipe}

\Copyright{Lu\'{\i}s Cruz-Filipe}

\subjclass{H.2.7 Database Administration, D.1.6 Logic Programming}

\keywords{grounded fixpoints, active integrity constraints}

\begin{document}

\maketitle

\begin{abstract}
  The formalism of active integrity constraints was introduced as a way to specify
  particular classes of integrity constraints over relational databases together with
  preferences on how to repair existing inconsistencies.
  The rule-based syntax of such integrity constraints also provides algorithms for finding
  such repairs that achieve the best asymptotic complexity.

  However, the different semantics that have been proposed for these integrity constraints
  all exhibit some counter-intuitive examples.
  In this work, we look at active integrity constraints using ideas from algebraic
  fixpoint theory.
  We show how database repairs can be modeled as fixpoints of particular operators on
  databases, and study how the notion of grounded fixpoint induces a corresponding notion
  of grounded database repair that captures several natural intuitions, and in particular
  avoids the problems of previous alternative semantics.

  In order to study grounded repairs in their full generality, we need to generalize the
  notion of grounded fixpoint to non-deterministic operators.
  We propose such a definition and illustrate its plausibility in the database context.
\end{abstract}

\section{Introduction}

The classical definition of model of a logic theory requires models to be deductively
closed.
An alternative phrasing of this fact is saying that models are fixpoints of some
entailment operator, and indeed the semantics of many modern logic frameworks can be
described as (minimal) fixpoints of particular operators -- in particular, those of logic
programs, default logics, or knowledge representation formalisms based on argumentation.

Several of these formalisms focus on models that can be constructed ``from the ground up''
(such as the minimal model of a positive logic program).
Grounded fixpoints of lattice operators, studied in~\cite{Bogaerts2015}, were proposed with
the intent of capturing this notion in the formal setting of algebraic fixpoint theory,
and were shown to abstract from many useful types of fixpoints in logic programming and
knowledge representation.

In this work, we are interested in applying this intuition within the context of
databases with integrity constraints -- formulas that describe logical relations between
data in a database, which should hold at all times.
We focus on the particular formalism of active integrity constraints (AICs), which not
only specify an integrity constraint, but also give indications on how inconsistent
databases can be repaired.
Although not all integrity constraints can be expressed in this formalism, AICs capture
the class of integrity constraints that can be written in denial clausal
form, which includes many examples that are important in practice~\cite{Flesca2004}.
Using AICs, one can distinguish between different types of repairs that embody typical
desirable properties -- minimality of change~\cite{Eiter1992,Winslett1990},
the common sense law of inertia~\cite{Przymusinski1997}, or non-circular justification for
repair actions~\cite{Caroprese2011}.
These intuitions capture many aspects of the idea of ``building a model from the ground
up'', present in grounded fixpoints.
However, the semantics of both founded~\cite{Caroprese2006} and justified
repairs~\cite{Caroprese2011} exhibit counter-intuitive behaviors, which led to the
proposal of well-founded repairs~\cite{CEGN2013}.
These in turn are not modular with respect to stratification of repairs~\cite{lcf:14},
rendering their computation problematic.

In this paper we show that repairs of inconsistent databases can be characterized as
fixpoints of a particular operator, with minimality of change corresponding to being a
minimal fixpoint, and that both founded and well-founded repairs can be described as
fixpoints with additional properties.
We then study grounded fixpoints of this operator, and show that they include all founded
and well-founded repair, but not all justified repairs.
In particular, grounded fixpoints avoid the circularity issues found in founded repairs,
while including some intuitive non-justified repairs.

To study at AICs in their full generality, we need to consider non-deterministic
operators.
While there is currently no notion of grounded fixpoint of a non-deterministic operator,
we show that we can define this concept in the context of AICs in a manner that naturally
generalizes the deterministic definition.
We then show how this in turn yields a plausible definition of grounded fixpoints of
non-deterministic operators within the general framework of algebraic fixpoint theory.

\subparagraph*{Related work.}
Database consistency has long since been recognized as an important problem in
knowledge management.
Especially in relational databases, integrity constraints have been used
for decades to formalize relationships between data in the database that are
dictated by its semantics~\cite{Abiteboul1995,Beeri1981}.

Whenever an integrity constraint is violated, it is necessary to change the database in
order to regain consistency.
This process of bringing the database back to consistency is known as
\emph{database repair}, and the problem of database repair is to determine whether such a
transformation is possible.
Typically, there are several possible ways of repairing an inconsistent database, and
several criteria have been proposed to evaluate them.
\emph{Minimality of change}~\cite{Eiter1992,Winslett1990} demands that the
database be changed as little as possible, while the
\emph{common-sense law of inertia}~\cite{Przymusinski1997} states that every change should
have an underlying reason.
While these criteria narrow down the possible database repairs, human interaction is
ultimately required to choose the ``best'' possible repair~\cite{Teniente1995}.

Database management systems typically implement integrity constraints as a variant of
event-condition-action rules (ECAs,~\cite{Teniente1995,Widom1996}), for which rule processing
algorithms have been proposed and a procedural semantics has been defined.
However, their lack of declarative semantics makes it difficult to understand the
behavior of multiple ECAs acting together and to evaluate rule-processing algorithms in a
principled way.
Active integrity constraints (AICs)~\cite{Flesca2004} are inspired by the same principle,
encoding an integrity constraint together with preferred update actions to repair it.
The update actions are limited to addition and removal of tuples from the database, as
this suffices to implement the three main operations identified in the seminal work of
Abiteboul~\cite{Abiteboul1988}.
AICs follow the tradition of expressing database dependencies through logic programming,
which is common namely in the setting of deductive
databases~\cite{Marek1995,Naqvi1988,Przymusinski1997}.

The declarative semantics for AICs~\cite{Caroprese2006,Caroprese2011} is based on the
concept of founded and justified repairs, motivated by different interpretations of the
common-sense law of inertia, and
the operational semantics for AICs~\cite{CEGN2013} allows their direct computation by
means of intuitive tree algorithms,
which have been implemented over SQL databases~\cite{KMIS2015}.
However, neither founded nor justified repairs are completely satisfactory, as
counter-intuitive examples have been produced exhibiting limitations of both types of
repairs.
Similar flaws have been exposed for the alternative notion of well-founded
repairs~\cite{CEGN2013}.

Deciding whether a database can be repaired is typically a computationally hard problem.
In the framework of AICs, the complexity of this problem depends on the type of repairs
allowed, varying between NP-complete and $\Sigma^2_p$.
Because of this intrinsic complexity, techniques to split a problem in several smaller
ones are important in practice.
A first step in this direction was taken in~\cite{Mayol1999}, but that work explicitly
forbids cyclic dependencies.
A more general study, in the context of AICs, was undertaken in~\cite{lcf:14}, which
studies conditions under which a set of constraints can be split into smaller sets,
whose repairs may then be computed separately.

In the more general setting of knowledge bases with more powerful reasoning abilities,
the problem of computing repairs is much more involved than in databases,
as it amounts to solving an abduction problem~\cite{Guessoum1998}.
In those frameworks, AICs can help greatly with finding repairs, and we are currently
investigating how this formalism can be applied outside the database world~\cite{CNS2016}.

The operational semantics for AICs proposed in~\cite{CEGN2013} was inspired by Antoniou's
survey on semantics of default logic~\cite{Antoniou1999}.
The realization that Reiter's original semantics for default logic~\cite{Reiter1980}
defines extensions by means of what is essentially a fixpoint definition naturally leads
to the question of whether we can characterize repairs of inconsistent databases in a
similar way.
Indeed, some connections between the semantics for AICs and logic programming have been
discussed in~\cite{Caroprese2011}, and fixpoints play a crucial role in defining several
semantics for logic programs~\cite{Fitting2002}.
These include the standard construction of minimal models of positive logic programs and
the notion of answer sets (via the Gelfond--Lifschitz transform).
Fixpoints also abound in other domains of logic; many of these occurrences of fixpoints
are summarized in~\cite{Bogaerts2015}, and showing that several of them can be seen as
instances of the same abstract notion constitutes one of those authors' motivation for
studying grounded fixpoints.

\section{Preliminaries}

In this section we review the concepts and results that are directly relevant for the
remainder of the presentation: grounded fixpoints of lattice
operators~\cite{Bogaerts2015}, the formalism of active integrity
constraints~\cite{Flesca2004}, founded~\cite{Caroprese2006},
justified~\cite{Caroprese2011} and well-founded~\cite{CEGN2013} (weak) repairs, and
parallelization results for these.

\subparagraph*{Grounded fixpoints.}
A partial order is a binary relation that is reflexive, antisymmetric and transitive.
A set $L$ equipped with a partial order $\leq$ is called a \emph{poset} (for partially
ordered set), and it is customary to write $x<y$ if $x,y\in L$ are such that $x\leq y$ and
$x\neq y$.
Given $S\subseteq L$, an \emph{upper bound} of $S$ is an element $x$ such that $s\leq x$
for all $s\in S$, and $x$ is a \emph{least upper bound} (lub) or \emph{join} of $S$ if
$x\leq y$ for all upper bounds $y$ of $S$, and we write $x=\bigvee S$.
The notion of \emph{(greatest) lower bound}, or \emph{meet}, is dually defined, and
written $\bigwedge S$.
Meets and joins, if they exist, are necessarily unique.
For binary sets, it is standard practice to write $x\wedge y$ and $x\vee y$ instead of
$\bigwedge\{x,y\}$ and $\bigvee\{x,y\}$.

A \emph{complete lattice} is a poset in which every set has a join and a meet.
In particular, complete lattices have a greatest element $\top$ and a smallest element
$\bot$.
The \emph{powerset lattice} of a set $S$ is $\langle\wp(S),\subseteq\rangle$, whose
elements are the subsets of $S$ ordered by inclusion.
The powerset lattice is a complete lattice with joins given by union and meets given by
intersection.
Its greatest element is $S$, and its smallest element is $\emptyset$.

A lattice operator is a function $\OO:L\to L$.
A \emph{fixpoint} of $\OO$ is an element $x\in L$ for which $\OO(x)=x$.
If $x\leq y$ for all fixpoints $y$ of $\OO$, then $x$ is said to be the \emph{least}
(or \emph{minimal} \emph{fixpoint} of $\OO$.
Lattice operators do not need to have fixpoints, but \emph{monotone} operators (i.e.~those
for which $x\leq y$ implies $\OO(x)\leq\OO(y)$) always have a minimal fixpoint.

We will be interested in two particular kinds of fixpoints, introduced
in~\cite{Bogaerts2015}.
We summarize the definitions and Propositions~3.3, 3.5 and~3.8 from that work.

\begin{definition}
  Let $\OO$ be an operator over a lattice $\langle L,\leq\rangle$.
  An element $x\in L$ is:
  \begin{itemize}
  \item \emph{grounded} for $\OO$ if $\OO(x\wedge v)\leq v$ implies $x\leq v$, for
    all $v\in L$;
  \item \emph{strictly grounded} for $\OO$ if there is no $y\in L$ such that $y<x$
    and $(\OO(y)\wedge x)\leq y$.
  \end{itemize}
\end{definition}

\begin{lemma}
  Let $\OO$ be an operator over a lattice $\langle L,\leq\rangle$.
  \begin{enumerate}
  \item All strictly grounded fixpoints of $\OO$ are grounded.
  \item If $\langle L,\leq\rangle$ is a powerset lattice, then all grounded fixpoints of
    $\OO$ are strictly grounded.
  \item All grounded fixpoints of $\OO$ are minimal.
  \end{enumerate}
\end{lemma}
We will be working mostly in a powerset lattice, so throughout this paper we will treat
the notions of strictly grounded and grounded as equivalent.

\subparagraph*{Active integrity constraints (AICs).}

The formalism of AICs was originally introduced in~\cite{Flesca2004}, but later simplified
in view of the results in~\cite{Caroprese2006}.
We follow the latter's definition, with a more friendly and simplified notation.

We assume a fixed set $\At$ of \emph{atoms} (typically, closed atomic formulas of a
first-order theory); subsets of $\At$ are \emph{databases}.
A \emph{literal} is either an atom ($a$) or its negation ($\neg a$), and a database $\DB$
satisfies a literal $\ell$, denoted $\DB\models\ell$, if: $\ell$ is an atom $a\in\DB$,
or $\ell$ is $\neg a$ and $a\not\in\DB$.
An \emph{update action} $\alpha$ has the form ${+a}$ or ${-a}$, where $a\in\At$; ${+a}$
and ${-a}$ are \emph{dual} actions, and we represent the dual of $\alpha$ by $\alpha^D$.
Update actions are intended to change the database: ${+a}$ adds $a$ to the database
(formally: it transforms $\DB$ into $\DB\cup\{a\}$), while ${-a}$ removes it (formally: it
transforms $\DB$ into $\DB\setminus\{a\}$).
A set of update actions $\U$ is \emph{consistent} if it does not contain an action and its
dual.
A consistent set of update actions $\U$ acts on a database $\DB$ by updating $\DB$ by
means of all its actions simultaneously; we denote the result of this operation by
$\U(\DB)$.

Literals and update actions are related by natural mappings $\lit$ and $\ua$, where
$\lit({+a})={a}$, $\lit({-a})={\neg a}$, $\ua(a)={+a}$ and $\ua({\neg a})={-a}$.
An AIC is a rule $r$ of the form
\begin{equation}
  \label{eq:AIC}
  \ell_1,\ldots,\ell_n\supset\alpha_1\mid\ldots\mid\alpha_k
\end{equation}
where $n,k\geq 1$ and
$\{\lit(\alpha_1^D),\ldots,\lit(\alpha_k^D)\}\subseteq\{\ell_1,\ldots,\ell_n\}$.
The intuition behind this notation is as follows: the \emph{body} of the rule,
$\body(r)=\ell_1,\ldots,\ell_n$ describes an inconsistent state of the database.
If $\DB\models\ell_1\wedge\ldots\wedge\ell_n$, which we write as $\DB\models\body(r)$,
then $r$ is \emph{applicable}, and we should fix this inconsistency by applying one of the
actions in the \emph{head} of $r$, $\head(r)=\alpha_1\mid\ldots\mid\alpha_k$.
The syntactic restriction was motivated by the observation~\cite{Caroprese2006} that
actions that do not satisfy this condition may be removed from $\head(r)$ without changing
the semantics of AICs, which we now describe.

Generic integrity constraints were previously written as first-order clauses with empty
head (see~\cite{Flesca2004}), and we can see AICs as a generalization of this concept: an
integrity constraint $\ell_1\wedge\ldots\wedge\ell_n\to\bot$ expresses no preferences
regarding repairs, and thus corresponds to the (closed instances of the) AIC
$\ell_1,\ldots,\ell_n\supset\ua(\ell_1)^D\mid\ldots\mid\ua(\ell_n)^D$.
Our presentation essentially treats $\At$ as a set of propositional symbols,
following~\cite{Caroprese2011}; for the purposes of this paper, the distinction is
immaterial (we can identify an AIC including variables with the set of its closed
instances), but our choice makes the presentation much simpler.

A set of update actions $\U$ is a \emph{weak repair} for $\DB$ and a set $\eta$ of AICs
(shortly, for $\Ieta$) if: (i)~every action in $\U$ changes $\DB$ and
(ii)~$\U(\DB)\not\models\body(r)$ for all $r\in\eta$.
Furthermore, if $\U$ is minimal wrt~set inclusion, then $\U$ is said to be a
\emph{repair}; repairs are also minimal among all sets satisfying only condition~(ii),
embodying the principle of
\emph{minimality of change}~\cite{Winslett1990} explained earlier.

\begin{definition}
  A set of update actions $\U$ is \emph{founded} wrt~$\Ieta$ if, for every
  $\alpha\in\U$, there exists $r\in\U$ such that $\alpha\in\head(r)$ and
  $\U(\DB)\models\body(r)\setminus\{\lit(\alpha^D)\}$.
  A \emph{founded (weak) repair} is a (weak) repair that is founded.
\end{definition}
The intuition is as follows: in a founded weak repair, every action has \emph{support} in
the form of a rule that ``requires'' its inclusion in $\U$.
We will use the (equivalent) characterization of founded sets: $\U$ is founded iff, for
every $\alpha\in\U$, there is a rule $r$ such that $\alpha\in\head(r)$ and
$(\U\setminus\{\alpha\})(\DB)\models\body(r)$.

However, Caroprese \emph{et al.}~\cite{Caroprese2011} discovered that there can be founded
repairs exhibiting \emph{circularity of support} (see Example~\ref{ex:founded} below), and
they proposed the stricter notion of justified repair.

\begin{definition}
  Let $\U$ be a set of update actions and $\DB$ be a database.
  \begin{itemize}
  \item The \emph{no-effect actions} wrt~$\DB$ and $\U$ are
    the actions that do not affect either $\DB$ or $\U(\DB)$:
    $\neff(\U)=\{{+a}\mid a\in\DB\cap\U(\DB)\}\cup\{{-a}\mid a\not\in\DB\cup\U(\DB)\}$.
  \item The set of \emph{non-updateable literals} of an AIC $r$ is
    $\body(r)\setminus\lit\left(\head(r)^D\right)$, where the functions $\lit$ and
    $\cdot^D$ are extended to sets in the natural way.
  \item $\U$ is \emph{closed under $\eta$} if, for each $r\in\eta$,
    $\ua(\nup(r))\subseteq\U$ implies $\head(r)\cap\U\neq\emptyset$.
  \item $\U$ is a \emph{justified action set} if it is the least superset of
    $\U\cup\neff(\U)$ closed under~$\eta$.
  \item $\U$ is a justified (weak) repair if $\U$ is a (weak) repair and $\U\cup\neff(\U)$
    is a justified action set.
  \end{itemize}
\end{definition}

The notion of justified weak repair, however, is extremely complicated and unwieldy in
practice, due to its quantification over sets of size comparable to that of $\DB$.
Furthermore, it excludes some repairs that seem quite reasonable and for which it can be
argued that the circularity of support they exhibit is much weaker (see
Example~\ref{ex:justified}).
This motivated proposing yet a third kind of weak repair: well-founded repairs, that are
defined by means of an operational semantics inspired by the syntax of
AICs~\cite{CEGN2013}.

\begin{definition}
  Let $\DB$ be a database and $\eta$ be a set of AICs.
  The \emph{well-founded repair tree} for $\Ieta$ is built as follows: its nodes are
  labeled by sets of update actions, with root $\emptyset$; the descendants of a node
  with consistent label $\U$ are all sets of the form
  $\U\cup\{\alpha\}$ such that there exists a rule $r\in\eta$ with $\alpha\in\head(r)$
  and $\U(\DB)\models\body(r)$.
  The consistent leaves of this tree are \emph{well-founded weak repairs} for
  $\Ieta$.
\end{definition}
Equivalently, a weak repair $\U$ for $\Ieta$ is well-founded iff there exists a sequence
of actions $\alpha_1,\ldots,\alpha_n$ such that $\U=\{\alpha_1,\ldots,\alpha_n\}$ and, for
each $1\leq i\leq n$, there exists a rule $r_i$ such that
$\{\alpha_1,\ldots,\alpha_{i-1}\}(\DB)\models\body(r_i)$ and $\alpha_i\in\head(r_i)$.

The availability of multiple actions in the heads of AICs makes the construction of
repairs non-deterministic, and a normalization procedure was therefore proposed
in~\cite{Caroprese2011}.
An AIC $r$ is \emph{normal} if $|\head(r)|=1$.
If $r$ is an AIC of the form in~\eqref{eq:AIC}, then
$\normalize(r)=\{\ell_1,\ldots,\ell_n\supset\alpha_i\mid 1\leq i\leq k\}$,
and $\normalize(\eta)=\bigcup\{\normalize(r)\mid r\in\eta\}$.
It is straightforward to check that $\U$ is a weak repair (respectively, repair, founded
(weak) repair or well-founded (weak) repair) for $\Ieta$ iff $\U$ is a weak repair
(resp.\ repair, founded (weak) repair or well-founded (weak) repair) for
$\langle\DB,\normalize(\eta)\rangle$; however, this equivalence does not hold for
justified (weak) repairs, as shown in~\cite{Caroprese2011}.

\subparagraph*{Parallelization.}

Determining whether a database satisfies a set of AICs is linear on both
the size of the database and the number of constraints.
However, determining whether an inconsistent database can be repaired is a much harder
problem -- NP-complete, if any repair is allowed, but $\Sigma^P_2$-complete,
when repairs have to be founded or justified.
(Here, $\Sigma^P_2$ is the class of problems that can be solved in non-deterministic polynomial
time, given an oracle that can solve any NP-complete problem.)
This complexity only depends on the size of the set of AICs~\cite{Caroprese2011}.
In the normalized case, several of these problems become NP-complete; even so, separating a set
of AICs into smaller sets that can be processed independently has a significant practical
impact~\cite{lcf:14}.

There are two important splitting techniques: \emph{parallelization}, which splits a set
of AICs into smaller sets for which the database can be repaired independently (in
principle, in parallel); and \emph{stratification}, which splits a set of AICs into
smaller sets, partially ordered, such that repairs can be computed incrementally using a
topological sort of the order.
We shortly summarize the definitions and results from~\cite{lcf:14}.

\begin{definition}
  Let $\eta_1$ and $\eta_2$ be two sets of AICs over a common set of atoms $\At$.
  \begin{itemize}
  \item $\eta_1$ and $\eta_2$ are \emph{strongly independent}, $\eta_1\indep\eta_2$, if,
    for each pair of rules $r_1\in\eta_1$ and $r_2\in\eta_2$, $\body(r_1)$ and
    $\body(r_2)$ contain no common or dual literals.
  \item $\eta_1$ and $\eta_2$ are \emph{independent}, $\eta_1\perp\eta_2$, if,
    for each pair of rules $r_1\in\eta_1$ and $r_2\in\eta_2$, $\lit(\head(r_i))$ and
    $\body(r_{3-i})$ contain no common or dual literals, for $i=1,2$.
  \item $\eta_1$ \emph{precedes} $\eta_2$, $\eta_1\prec\eta_2$, if, for each pair of
    rules $r_1\in\eta_1$ and $r_2\in\eta_2$, $\lit(\head(r_2))$ and $\body(r_1)$ contain
    no common or dual literals, but not conversely.
\end{itemize}
\end{definition}
From the syntactic restrictions on AICs, it follows that $\eta_1\indep\eta_2$ implies
$\eta_1\perp\eta_2$.
Given two sets of AICs $\eta_1$ and $\eta_2$ a set of update actions $\U$, let
$\U_i=\U\cap\{\alpha\mid\alpha\in\head(r),r\in\eta_i\}$.
\begin{lemma}
  \label{lem:foiks}
  Let $\eta_1$ and $\eta_2$ be sets of AICs, $\eta=\eta_1\cup\eta_2$, and $\U$ be a
  set of update actions.
  \begin{enumerate}
  \item If $\eta_1\indep\eta_2$, then $\U$ is a repair for $\Ieta$ iff $\U=\U_1\cup\U_2$
    and $\U_i$ is a repair for $\Ieta[i]$, for $i=1,2$.
  \item If $\eta_1\perp\eta_2$, then $\U$ is a founded/well-founded/justified repair for
    $\Ieta$ iff $\U=\U_1\cup\U_2$ and $\U_i$ is a founded/well-founded/justified repair
    for $\Ieta[i]$, for $i=1,2$.
  \item If $\eta_1\prec\eta_2$, then $\U$ is a founded/justified repair for $\Ieta$ iff
    $\U=\U_1\cup\U_2$, $\U_1$ is a founded/justified repair for $\Ieta[1]$ and $\U_2$
    is a founded/justified repair for $\langle\U_1(\DB),\eta_2\rangle$.
  \end{enumerate}
\end{lemma}

\section{Repairs as Fixpoints}

In this section we show how a set of AICs induces an operator on a suitably defined lattice.
This operator is in general non-deterministic; in order to reuse the results from
algebraic fixpoint theory, we restrict our attention to the case of normalized AICs, and
delay the discussion of the general case to a later section.

\subparagraph*{The operator $\T$.}
Throughout this paragraph, we assume $\DB$ to be a fixed database over a set of atoms $\At$
and $\eta$ to be a set of AICs over $\At$.

The intuitive reading of an AIC $r$ naturally suggests an operation on sets of update
actions $\U$, defined as ``if $\U(\DB)\models\body(r)$ holds, then add $\head(r)$ to
$\U$''.
However, this definition quickly leads to inconsistent sets of update actions, which we
want to avoid.
We therefore propose a slight variant of this intuition.

\begin{definition}
  Let $\U$ and $\V$ be consistent sets of update actions over $\At$.
  The set $\U\uplus\V$ is defined as
  $(\U\cup\{\alpha\in\V\mid\alpha^D\not\in\U\})\setminus\{\alpha\in\U\mid\alpha^D\in\V\}$.
\end{definition}
This operation models sequential composition of repairs in the following sense: if every
action in $\U$ changes $\DB$ and every action in $\V$ changes $\U(\DB)$, then
$(\U\uplus\V)(\DB)=\V(\U(\DB))$.
Furthermore, if $\U$ and $\V$ are both consistent, then so is $\U\uplus\V$.

We can identify subsets of $\At$ with sets of update actions by matching each atom $a$
with the corresponding action that changes the database (i.e.~$-a$ if $a\in\DB$ and $+a$
otherwize).
We will abuse notation and use this bijection implicitly, so that we can reason over the
powerset lattice $\langle\wp(\At),\subseteq\rangle$ as having sets of update actions as
elements.

\begin{definition}
  The operator $\TT:\wp(\At)\to\wp(\wp(\At))$ is defined as follows:
  $\U\uplus\V\in\TT(\U)$ iff $\V$ can be constructed by picking exactly one action from
  the head of each rule $r$ such that $\U(\DB)\models\body(r)$.
\end{definition}
Each set $\V$ may contain less update actions than there are rules $r$ for which
$\U(\DB)\models\body(r)$, as the same action may be chosen from the heads of different
rules; and there may be rules $r$ for which $|\head(r)\cap\V|>1$.
This is illustrated in the following simple example.
\begin{example}
  Let $\DB=\{a,b\}$ and $\eta=\{a,b,\neg c\supset{-a}\mid{-b};\quad a,b,\neg d\supset{-a}\mid{-b}\}$.
  Then $\TT(\emptyset)=\{\{{-a}\},\{{-b}\},\{{-a},{-b}\}\}$: the bodies of both rules are
  satisfied in $\DB$, and we can choose $-a$ from the heads of both, $-b$ from the heads
  of both, or $-a$ from one and $-b$ from the other.
\end{example}

The syntactic restrictions on AICs guarantee that all sets $\V$ in the above definition
are consistent: if $+a,-a^D\in\V$, then there are rules $r_1$ and $r_2$ such that
$\neg a\in\body(r_1)$ and $a\in\body(r_2)$ with $\U(\DB)\models\body(r_i)$ for
$i=1,2$, which is impossible.
In the interest of legibility, we will write $\T$ instead of $\TT$ whenever $\DB$ and
$\eta$ are clear from the context.

\subparagraph*{The normalized case.}
In the case that $\eta$ contains only normalized AICs, the set $\T(\U)$ is a singleton,
and we can see $\T$ as a lattice operator over $\langle\wp(\At),\subseteq\rangle$.
We will assume this to be the case throughout the remainder of this section, and by abuse
of notation use $\T$ also in this situation.
In the normalized case, we thus have
\[\T(\U) = \U\uplus\{\head(r)\mid\U(\DB)\models\body(r)\}\,.\]

Since we can always transform $\eta$ in a set of normalized AICs by the transformation
$\normalize$ defined above, in most cases it actually suffices to consider this simpler
scenario, which warrants its study.
The exception is the case of justified repairs for non-normalized AICs, which we defer to
a later section.
All our results also apply to general integrity constraints by seeing them as AICs with
maximal heads and applying $\normalize$ to the result.

The operator $\T$ characterizes the notions of weak repair, repair, founded and
well-founded sets of update actions.

\begin{lemma}
  \label{lem:weak-repair}
  $U$ is a weak repair for $\Ieta$ iff $\U$ is a fixpoint of
  $\T$.
\end{lemma}

\begin{lemma}
  \label{lem:repair}
  $U$ is a repair for $\Ieta$ iff $\U$ is a minimal fixpoint of
  $\T$.
\end{lemma}

\begin{lemma}
  \label{lem:founded-char}
  A consistent set of update actions $\U$ is founded wrt~$\Ieta$ iff, for all
  $\alpha\in\U$, it is the case that $\alpha\in\T(\U\setminus\{\alpha\})$.
\end{lemma}

\begin{lemma}
  \label{lem:wf}
  A weak repair $\U$ for $\Ieta$ is well-founded iff there is an ordering
  $\alpha_1,\ldots,\alpha_n$ of the elements of $\U$ such that
  $\alpha_i\in\T(\{\alpha_1,\ldots,\alpha_{i-1}\})$ for each $i=1,\ldots,n$.
\end{lemma}

The correspondence between justified repairs and answer sets for particular logic
programs~\cite{Caroprese2011} shows that justified repairs can also be characterized in a
related manner.
However, since answer sets of a logic program are models of its Gelfond--Lifschitz
transform, the corresponding characterization in terms would be as fixpoints of the
corresponding operator for a similarly derived set of AICs, rather than of $\T$.
This characteristic of justified repairs also explains the rather unexpected behavior we
will see later, in \S~\ref{sec:general}.

\subparagraph*{Grounded fixpoints of $\T$.}
Founded, well-founded and justified repairs were all introduced with the purpose of
characterizing a class of repairs whose actions are supported (there is a reason for
having them in the set), and that support is not circular; in particular, these repairs
should be constructible ``from the ground up'', which was the motivation for defining
well-founded repairs.
However, all notions exhibit unsatisfactory examples: there exist founded repairs with
circular support~\cite{Caroprese2011} and repairs with no circular support that
are not justified~\cite{CEGN2013}; well-founded repairs, on the other hand, are not
stratifiable~\cite{lcf:14}, which impacts their computation in practice.

Following the intuition in~\cite{Bogaerts2015} that grounded fixpoints capture the idea of
building fixpoints ``from the ground up'', we propose the following notion of $\T$.

\begin{definition}
  A repair $\U$ for $\Ieta$ is \emph{grounded} if $\U$ is a grounded fixpoint
  of $\T$.
\end{definition}

Since we are working within a powerset lattice, the notions of grounded and strictly
grounded fixpoints coincide.
As it turns out, the latter notion is most convenient for the proofs of our results.
We thus characterize grounded repairs as repairs $\U$ such that: if
$\V\subsetneq\U$, then $\T(\V)\cap\U\not\subseteq\V$.
Equivalently: if $\V\subsetneq\U$, then
$\T(\V)\cap(\U\setminus\V)\neq\emptyset$.

Since all grounded fixpoints are minimal, it makes no sense to define grounded weak
repairs.
The notion of grounded fixpoint therefore intrinsically embodies the principle of
minimality of change, unlike other kinds of weak repairs previously defined.
Furthermore, grounded repairs also embody the notion of ``support'' previously defined.

\begin{lemma}
  \label{lem:founded}
  Every grounded repair for $\Ieta$ is both founded and well-founded.
\end{lemma}

However, the notion of grounded repair is strictly stronger than both of these:
the first example, from~\cite{CEGN2013}, also shows that some forms of circular
justifications are avoided by grounded repairs.

\begin{example}
  \label{ex:founded}
  Let $\DB=\{a,b\}$ and $\eta=\{
  a,\neg b \supset{-a};\quad
  a,\neg c \supset{+c};\quad
  \neg a,b \supset{-b};\quad
  b,\neg c \supset{+c}\}$.
  Then $\U=\{-a,-b\}$ is a founded repair that is not grounded: $\V=\emptyset$
  satisfies $\T(\V)\cap\U=\{+c\}\cap\U=\emptyset\subseteq\V$.
  The more natural repair $\U'=\{+c\}$ is both founded and grounded.
\end{example}

\begin{example}
  \label{ex:well-founded}
  Let $\DB=\emptyset$ and $\eta=\{
  a,\neg b,\neg c \supset{+c};\quad
  \neg a,\neg b \supset{+b};\quad
  \neg a \supset{+a}\}$.
  There are two well-founded repairs for $\Ieta$: $\U_1=\{{+a},{+c}\}$ (obtained by applying
  the last rule and then the first) and $\U_2=\{{+b},{+a}\}$ (obtained by applying the second
  rule and then the last).
  However, $\U_2$ is not founded ($+b$ is not founded), so it cannot be grounded:
  indeed, $\V=\{{+a}\}$ is a strict subset of $\U_2$, and
  $\T(\V)\cap\U=\{{+a},{+b}\}\cap\U=\emptyset\subseteq\V$.
\end{example}
Also in this last example the grounded repair ($\U_1$) is somewhat more natural.

We now investigate the relation to justified repairs, and find that all justified repairs
are grounded, but not conversely -- confirming our earlier claim that the notion of
justified repair is too strong.

\begin{lemma}
  \label{lem:justified}
  Every justified repair for $\Ieta$ is grounded.
\end{lemma}

This result is not very surprising: justified weak repairs are answer sets of a
particular logic program (Theorem 6 in~\cite{Caroprese2011}), and in turn answer sets of
logic programs are grounded fixpoints of the consequence operator (see remark at the top
of \S~5 in~\cite{Bogaerts2015}).
However, the translation defined in~\cite{Caroprese2011} is from logic programs to
databases with AICs (rather than the other way around), so Lemma~\ref{lem:justified} is
\emph{not} a direct consequence of those results.

The notion of justified repair is also stricter than that of grounded repair, as the
following example from~\cite{Caroprese2011} shows.
\begin{example}
  \label{ex:justified}
  Let $\DB=\{a,b\}$ and $\eta=\{
  a,b \supset{-a};\quad
  a,\neg b \supset{-a};\quad
  \neg a,b \supset{-b}\}$.
  Then $\U=\{-a,-b\}$ is not justified (see~\cite{Caroprese2011}), but it is
  grounded: if ${-a}\in\V\subsetneq\U$, then $\T(\V)\cap\U$ contains
  ${-b}\in\U\setminus\V$, else $\T(\V)\cap\U$ contains ${-a}\in\U\setminus\V$.
\end{example}
This example was used in~\cite{CEGN2013} to point out that justified repairs sometimes
eliminate ``natural'' repairs; in this case, the first rule clearly motivates the action
$-a$, and the last rule then requires $-b$.
This is in contrast to Example~\ref{ex:founded}, where there was no clear reason to
include either $-a$ or $-b$ in a repair.
So grounded repairs avoid this type of unreasonable circularities, without being
as restrictive as justified repairs.

We thus have that grounded repairs are always founded and well-founded; the next example
shows that they do not correspond to the intersection of those classes.

\begin{example}
  Assume that $\DB=\emptyset$ and $\eta$ contains the following integrity
  constraints.
  \begin{align*}
    \neg a,\neg b&\supset{+a} &
    a,\neg b&\supset{+b} &
    \neg a,b&\supset{-b} &
    a,b,\neg c&\supset{+c} &
    a,\neg b,c&\supset{+b} &
    \neg a,b,c&\supset{+a}
  \end{align*}
  Then $\U=\{{+a},{+b},{+c}\}$ is a repair for $\Ieta$:
  the first three constraints require
  ${+a}$ and ${+b}$ to be included in any repair for $\Ieta$, and
  the last three state
  that no $2$-element subset of $\U$ is a repair.
  Furthermore, $\U$ is founded
  (the three last rules ensure that)
  and well-founded (starting with $\U$, the rules force us to add ${+a}$, ${+b}$ and ${+c}$,
  in that order).

  However, $\U$ is not strictly grounded for $\T$: if $\V=\{{+b}\}$, then
  $\V\subsetneq\U$, but $\T(\V)\cap\U=\emptyset\cap\U=\emptyset\subseteq\V$.
\end{example}
In this situation, $\U$ actually seems reasonable; however, observe that the support for
its actions \emph{is} circular: it is the three rules in the second row that make $\U$
founded, and none of them is applicable to $\DB$.
Also note that $\V(\DB)$ is a database for which the given set $\eta$ behaves very
awkwardly: the only applicable AIC tells us to remove $b$, but the only possible repair
is actually $\{{+a},{+c}\}$.

We do not feel that this example weakens the case for studying ground repairs, though: the
consensual approach to different notions of repair is that they express
\emph{preferences}.
In this case, where $\Ieta$ admits no grounded repair, it is sensible to allow a repair in
a larger class -- and a repair that is both founded and well-founded is a good candidate.
The discussion in \S~8 of~\cite{Caroprese2011} already proposes such a ``methodology'':
choose a repair from the most restrictive category (justified, founded, or any).
We advocate a similar approach, but dropping justified repairs in favor of grounded
repairs, and preferring well-founded to founded repairs.

The relations between the different classes of repairs are summarized in the picture below.

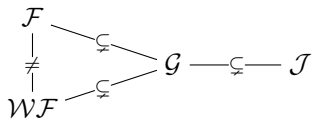
\[\xymatrix@R-2em@C+1em{%
  \mathcal F \ar@{-}[dd]|{\neq} \ar@{-}[dr]|{\subsetneq} \\
  & \mathcal G \ar@{-}[r]|{\subsetneq} & \mathcal J \\
  \mathcal{WF} \ar@{-}[ur]|{\subsetneq}
}
\]

We conclude this section with a note on complexity.

\begin{theorem}
  \label{thm:grounded-complexity}
  The problem of deciding whether there exist grounded repairs for $\Ieta$ is
  $\Sigma^P_2$-complete.
\end{theorem}

This result still holds if we allow a truly first-order syntax for AICs, where the atoms
can include variables that are implictly universally quantified.

\section{Parallelism}\label{sec:par}

Lemma~\ref{lem:foiks} shows that splitting a set of AICs into smaller ones transforms
the problem of deciding whether an inconsistent database can be repaired (and computing
founded or justified repairs) into smaller ones, with important practical
consequences.
The goal of this section is to show that grounded repairs enjoy similar properties.
This is even more relevant, as deciding whether grounded repairs exist is
presumably\footnote{I.e., assuming that $\mbox{P}\neq\mbox{NP}$.} more complex than for
the other cases, in view of Theorem~\ref{thm:grounded-complexity}.
For parallelization, we will go one step further, and propose a lattice-theoretical concept
of splitting an operator into ``independent'' operators in such a way that strictly
grounded fixpoints can be computed in parallel.

We make some notational conventions for the remainder of this section.
We will assume as before a fixed database $\DB$ and set of AICs $\eta$ over the same set
of atoms $\At$.
Furthermore, we will take $\eta_1$ and $\eta_2$ to be disjoint sets with
$\eta=\eta_1\cup\eta_2$, and write $\T_i$ for $\T^\DB_{\eta_i}$.
Also, we write $\ii$ for $3-i$ (so $\ii=1$ if $i=2$ and vice-versa).

\subparagraph*{Independence.}
We begin with a simple consequence of independence.

\begin{lemma}
  \label{lem:indep-com}
  If $\eta_1\perp\eta_2$, then $\T_1$ and $\T_2$ commute and
  $\T=\T_1\circ\T_2=\T_2\circ\T_1$.
\end{lemma}

The converse is not true.
\begin{example}
  Let $\eta_1=\{a,b\supset{-b}\}$ and $\eta_2=\{\neg a,\neg b\supset{+b}\}$.
  Then $\eta_1\not\perp\eta_2$, but $\T_1$ and $\T_2$ commute: if $a\in\U(\DB)$, then
  $\T_1(\T_2(\U))=\T_1(\U)=\T_2(\T_1(\U))$; otherwise,
  $\T_1(\T_2(\U))=\T_2(\U)=\T_2(\T_1(\U))$.
\end{example}

\begin{lemma}
  \label{lem:grounded-par}
  A set of update actions $\U$ is a grounded repair for $\Ieta$ iff $\U=\U_1\cup\U_2$ and
  $\U_1$ is a grounded repair for $\Ieta[1]$ and $\U_2$ is a grounded repair for
  $\Ieta[2]$.
\end{lemma}

These properties are actually not specific to operators induced by AICs, but can be
formulated in a more general lattice-theoretic setting.

\begin{definition}
  Let $\langle L,\leq\rangle$ be a complete distributive lattice with complements.
  An operator $\OO:L\to L$ is an $(u,v)$-operator, with $u\leq v\in L$, if, for every
  $x\in L$,
  \[\OO(x)=\left(\OO(x\wedge v)\wedge u\right)\vee(x\wedge\bar u)\,.\]
\end{definition}
Intuitively, an $(u,v)$-operator only depends on the ``$v$-part'' of its argument, and the
result only differs from the input in its ``$u$-part''.
In this context, Proposition~3.5 of~\cite{Bogaerts2015} applies, so grounded and strictly
grounded fixpoints coincide; furthermore, we can extend the definition of independence to
this setting and generalize Lemmas~\ref{lem:indep-com} and~\ref{lem:grounded-par}.

Observe that, by construction, $\T_\eta$ is a $(\U,\V)$-operator with
$\U=\{\head(r)\mid r\in\eta\}$ and $\V=\{\ua(l)\mid l\in\body(r),r\in\eta\}$.

\begin{definition}
  Two operators $\OO_1,\OO_2:L\to L$ are \emph{independent} if each $\OO_i$ is an
  $(u_i,v_i)$-operator with $u_i\wedge v_{\ii}=\bot$.
\end{definition}

\begin{lemma}
  \label{lem:lattice-indep}
  If $\OO_1$ and $\OO_2$ are independent, then $\OO_1$ and $\OO_2$ commute.
  In this case, if $\OO$ is their composition, then $x\in L$ is (strictly) grounded for
  $\OO$ iff $x=(x\wedge v_1)\vee(x\wedge v_2)$ and $x\wedge v_i$ is (strictly) grounded
  for $\OO_i$.
\end{lemma}

This provides an algebraic counterpart to the parallelization of AICs, albeit requiring
that the underlying lattice be distributive and complemented: we say that $\OO$ is
parallelizable if there exist $\OO_1$ and $\OO_2$ in the conditions of
Lemma~\ref{lem:lattice-indep}, with $\OO=\OO_1\circ\OO_2$.
As in the original work~\cite{lcf:14}, it is straightforward to generalize these results to
finite sets of independent operators.

\subparagraph*{Stratification.}
We now consider the case where $\eta_1$ and $\eta_2$ are not independent, but rather
stratified, and show that part 3 of Lemma~\ref{lem:foiks} also applies to grounded repairs.

\begin{lemma}
  \label{lem:grounded-strat}
  Suppose that $\eta_1\prec\eta_2$.
  Then $\U$ is a grounded repair for $\Ieta$ iff $\U=\U_1\cup\U_2$, $\U_1$ is a grounded
  repair for $\Ieta[1]$, and $\U_2$ is a grounded repair for
  $\langle\U_1(\DB),\U_2\rangle$.
\end{lemma}

Unlike parallelization, there is no clear generalization of these results to a more
general setting: the definition of $\T_2$ is dependent of the particular fixpoint for
$\T_1$, and to express this dependency we are using the sets $\eta_1$ and $\eta_2$ in an
essential way.

\section{General AICs and Non-deterministic Operators}\label{sec:general}

We now return to the original question of defining grounded repairs for databases with
arbitrary (not necessarily normal) sets of active integrity constraints.
This requires generalizing the definition of (strictly) grounded element to
non-deterministic lattice operators, a question that was left open in~\cite{Bogaerts2015}.
We propose possible definitions for these concepts, and show that they exhibit desirable
properties in our topic of interest.

Let $\OO:L\to L$ be a lattice operator, and define its non-deterministic counterpart
$\nondet\OO:L\to\wp(L)$ by $\nondet\OO(x)=\{\OO(x)\}$.
A reasonable requirement is that $x$ should be (strictly) grounded for $\nondet\OO$ iff
$x$ is (strictly) grounded for $\OO$.
Furthermore, in the case of AICs we can also define a converse transformation: since every
set of AICs $\eta$ can be transformed into a normalized set $\neta$, we will
also require that $\U$ be a grounded repair for $\T_\eta$ iff $\U$ is a grounded repair
for $\T_\neta$.

\begin{definition}
  Let $\OO:L\to\wp(L)$ be a non-deterministic operator over a complete lattice $L$.
  An element $x\in L$ is:
  \begin{itemize}
  \item \emph{grounded} for $\OO$ if $\left(\bigvee\OO(x\wedge v)\right)\leq v$
    implies $x\leq v$;
  \item \emph{strictly grounded} for $\OO$ if there is no $v<x$ such that
    $\left(\bigvee\OO(v)\right)\wedge x\leq v$.
  \end{itemize}
\end{definition}
Clearly these definitions satisfy the first criterion stated above: given $\OO:L\to L$,
$\bigvee(\nondet\OO(x))=\OO(x)$ for every $x\in L$.
The choice of a join instead of a meet is motivated by the second criterion, which we will
show is satisfied by this definition.
Furthermore, all grounded elements are again strictly grounded, and the
two notions coincide over powerset lattices -- the proofs
in~\cite{Bogaerts2015} are trivial to adapt to this case.

As before, we assume that the database $\DB$ is fixed, and omit it from the superscript in
the operators below.

\begin{lemma}
  \label{lem:nondet-subset}
  For every $\U$, $\T_{\neta}(\U)\subseteq\bigcup\T_\eta(\U)$.
\end{lemma}
Note that the set $\left\{\bigcup\head(r)\mid\U(\DB)\models\body(r),r\in\eta\right\}$ is
consistent, due to the syntactic restrictions on AICs and the fact that all rules are
evaluated in the same context.

\begin{example}
  The inclusion in Lemma~\ref{lem:nondet-subset} is, in general, strict: consider
  $\DB=\emptyset$, $\U=\{+a\}$, and let $\eta=\{a,\neg b\supset{-a}\mid{+b}\}$.
  Then $\neta$ contains the two AICs $a,\neg b\supset{-a}$ and $a,\neg b\supset{+b}$.
  In this case, $\T_\eta(\U)=\{\emptyset,\{{+a},{+b}\}\}$ and $\T_\neta(\U)=\{{+b}\}$.
\end{example}

\begin{lemma}
  \label{lem:nondet-grounded}
  $\U$ is strictly grounded for $\T_\eta$ iff $\U$ is strictly grounded for $\T_\neta$.
\end{lemma}

A fixpoint of a non-deterministic operator $\OO:L\to\wp(L)$ is a value $x\in L$ such that
$x\in\OO(L)$ (see e.g.~\cite{Khamsi1993}).
From the definition of $\T_\eta$, it is immediate that $\U\in\T_\eta(\U)$ iff
$\T(\U)=\{\U\}$.
Furthermore, Lemmas~\ref{lem:weak-repair} and~\ref{lem:repair} still hold in this
non-deterministic case, allowing us to derive the following consequence of the previous
lemma.

\begin{corollary}
  \label{cor:nondet-grounded}
  $\U$ is a grounded repair for $\Ieta$ iff $\U$ is a grounded repair for
  $\langle\DB,\neta\rangle$.
\end{corollary}

Since repairs, founded repairs and well-founded repairs for $\eta$ and for $\neta$ also
coincide, we immediately obtain generalizations of Lemma~\ref{lem:founded}
for the general setting, and the parallelization and
independence results from \S~\ref{sec:par} also apply.

As observed in~\cite{Caroprese2011}, normalization does not preserve justified repairs.
Therefore, Lemma~\ref{lem:justified} does not guarantee that justified
repairs are always grounded in the general case.
Indeed, the next example shows that this is \emph{not} true.

\begin{example}
  Let $\DB=\emptyset$ and take $\eta$ to be the following set of AICs.
  \begin{align*}
         a,     b,\neg c &\supset{-a}\mid{-b}\mid{+c} &(1)&&
         a,\neg b        &\supset{-a}                 &(3)&&
    \neg a,     b,     c &\supset{+a}\mid{-b}\mid{-c} &(5)\\
    \neg a,     b,\neg c &\supset{+a}\mid{-b}\mid{+c} &(2)&&
         a,\neg b,     c &\supset{+b}\mid{-c}         &(4)&&
    \neg a,\neg b,     c &\supset{+a}\mid{+b}\mid{-c} &(6)\\
    &&&&&&&&
    \neg a,\neg b,\neg c &\supset{+a}\mid{+b}\mid{+c} &(7)
  \end{align*}
  Then $\U=\{{+a},{+b},{+c}\}$ is the only repair for $\Ieta$, and it is justified.
  Indeed, if $\V\subseteq\U$ is such that $\V\cup\neff(\U)$ is closed under $\eta$, then
  $\V$ must contain an action in the head of each of rules~$(1)$, $(2)$, $(5)$, $(6)$
  and~$(7)$.
  Since $\V\subseteq\U$, it follows that ${+c}\in\V$ (by~$(1)$) and that ${+a}\in\V$
  (by~$(5)$).
  But then $\V$ contains the actions corresponding to the non-updatable literals in
  rule~$(4)$ (namely, $+a$), and hence also ${+b}\in\V$, so $\V=\U$.

  However, $\U$ is not a strictly grounded fixpoint of $\T$: taking $\V=\{{+a}\}$, we see
  that the only rule applicable in $\V(\DB)$ is rule~$(3)$, and thus
  $\T(\V)=\{\emptyset\}$, from which trivially
  $\left(\bigcup\T(\V)\right)\cap\U\subseteq\V$.
\end{example}

An examination of the conditions under which $\Ieta$ may admit a justified repair that is
not strictly grounded shows that this example is among the simplest possible.
It is important to point out that $\U$ is also not a justified repair for
$\langle\DB,\normalize(\eta)\rangle$, either, which seems to suggest that origin of the
problem lies in the standard interpretation of AICs with non-singleton heads.
We plan to look further into the semantics of repairs for non-normal AICs in future work.

\section{Conclusions and Future Work}

We have presented a formalization of the theory of active integrity constraints in lattice
theory, by showing how a set of AICs $\eta$ over a database $\DB$ induces an operator
$\T^\DB_\eta$ over a suitably defined lattice of database repairs.
We characterized the standard notions of (weak) repairs, founded and well-founded repairs
in terms of this operator.
By studying the grounded fixpoints of $\T^\DB_\eta$ in the normalized case, we showed that
we obtain a notion of repair that is stricter than founded or well-founded repairs, but
more general than the problematic notion of justified repairs.
Furthermore, by suitably extending the notions of grounded and strictly grounded fixpoint
of a lattice operator to the non-deterministic case, we gained a general notion of
grounded repair also in the non-normalized case.
We also showed that grounded repairs are preserved under normalization, and that they
share the parallelization and stratification properties of founded and justified repairs
that are important for their practical applications.

Conversely, we were able to state some of the results in the database setting more
generally.
Thus, not only did we propose an extension of the notion of (strictly) grounded fixpoint
to the case of non-deterministic lattice operators, but we also defined what it means
for an operator to be parallelizable, and showed that several properties of
parallelizable operators are not specific to the database case.

We believe the concept of grounded repair to be the one that better captures our
intuitions on what a ``good'' repair is, in the framework of AICs.
We plan to use this notion as the basis for future work on this topic, namely concerning
the extension of AICs to more general knowledge representation formalisms, following the
proposals in~\cite{CNS2016}.

\bibliographystyle{plain}
\bibliography{cruzfilipe}

\appendix

\appendix

\section{}

In this appendix, we include the proofs of the results in the body of the paper.

\begin{proof}[Lemma~\ref{lem:weak-repair}]
  If $\U$ is a weak repair for $\Ieta$, then $\U(\DB)\not\models\body(r)$ for all $r\in\eta$,
  whence $\T(\U)=\U$.
  If $\U$ is not a weak repair for $\Ieta$, then $\U(\DB)\models\body(r)$ for some $r\in\eta$,
  and $\T(\U)$ differs from $\U$ in (at least) $\head(r)$.
\end{proof}

\begin{proof}[Lemma~\ref{lem:repair}]
  Direct from Lemma~\ref{lem:weak-repair} and the definition of repair.
\end{proof}

\begin{proof}[Lemma~\ref{lem:founded-char}]
  An action $\alpha\in\U$ is founded iff there is a rule $r\in\eta$ such that
  $\U(\DB)\models\body(r)\setminus\{\alpha^D\}$.
  This is equivalent to saying that $(\U\setminus\{\alpha\})(\DB)\models\body(r)$.
  But, by definition of $\T$, we have $\alpha\in\T(\U\setminus\{\alpha\})$ iff there is a
  rule $r\in\eta$ such that $(\U\setminus\{\alpha\})(\DB)\models\body(r)$, which
  concludes the proof.
  Consistency of $\U$ is needed for the direct implication, as $\alpha$ is only added to
  $\U\setminus\{\alpha\}$ by $\T$ if that set does not already contain $\alpha^D$.
\end{proof} 

\begin{proof}[Lemma~\ref{lem:wf}]
  If $\U$ is well-founded, then the ordering is given by the sequence of actions
  introduced at each node in the path, in the well-founded repair tree for $\Ieta$, going
  from the root to the node with label $\U$.
  Conversely, if $\U$ can be obtained in the manner described, then it defines a valid
  path in that same tree ending at a leaf.
\end{proof}

\begin{proof}[Lemma~\ref{lem:founded}]
  \begin{itemize}
  \item Assume that $\U$ is a grounded repair for $\Ieta$.
  For each $\alpha\in\U$, necessarily $\T(\U\setminus\{\alpha\})\cap\U\subsetneq(\U\setminus\{\alpha\})$,
  which implies that $\alpha\in\T(\U\setminus\{\alpha\})$.
  By Lemma~\ref{lem:founded-char}, this implies that $\U$ is founded.

  \item Let $\U$ be a grounded repair for $\Ieta$.
  Construct the sequence for well-founded repairs always choosing $u_i\in\U$ until this
  is no longer possible.
  Letting $\U'$ be the last constructed set, by construction both $\U'\subseteq\U$ and
  $\T(\U')\cap\U\subseteq\U'$ (otherwise we could proceed).
  Since $\U$ is grounded, it cannot be the case that $\U'\subsetneq\U$, so $\U=\U'$ and it
  is thus a well-founded repair.
  \end{itemize}
\end{proof}

\begin{proof}[Lemma~\ref{lem:justified}]
  Let $\U$ be a justified repair for $\Ieta$ and assume that $\V\subsetneq\U$.
  Then $\V\cup\neff(\U)$ is not closed under $\eta$, whence there is a rule $r\in\eta$
  such that $\ua(\nup(r))\subseteq\V\cup\neff(\U)$ and
  $\head(r)\not\in\V\cup\neff(\U)$.

  Since $V\subseteq\U$, also $\ua(\nup(r))\subseteq\U\cup\neff(\U)$, whence
  $\head(r)\in\U\cup\neff(\U)$ as $\U$ is closed under $\eta$.
  But $\head(r)\not\in\V\cup\neff(\U)$, so $\head(r)\in\U\setminus\V$.

  Then $\V(\DB)\models\body(r)$: on the one hand, $\ua(\nup(r))\subseteq\V\cup\neff(\U)$
  implies that $\V(\DB)\models\nup(r)$, as $\neff(\U)\subseteq\neff(\V)$; on the other
  hand, from $\head(r)\in\U$ we know that $\lit(\head(r))^D\in\DB$ (all actions in
  $\U$ change $\DB$), whence $\V(\DB)\models\lit(\head(r)^D)$ since $\head(r)\not\in\V$.
  As $r$ is normalized, there are no more literals in $\body(r)$, so
  $\V(\DB)\models\body(r)$ and therefore $\head(r)\in\T(\V)$.

  We thus conclude that $\T(\V)\cap\U\not\subseteq\V$.
  By arbitrariness of $\V$, it follows that $\U$ is grounded.
\end{proof}

\begin{proof}[Theorem~\ref{thm:grounded-complexity}]
  For membership, we need to show that we can decide the problem with a non-deterministic
  Turing machine with an NP oracle.
  Given a set of update actions $\U$, checking that it is a fixpoint of $\T$ can be done
  in polynomial time on the size of $\DB$ and $\eta$; the NP-oracle can then answer
  whether there exists $\V\subsetneq\U$ with $\T(\V)\cap\U\subseteq\V$, thereby
  establishing whether $\U$ is grounded.

  For hardness, we invoke the (polynomial time) translation $\mathit{aic}$ from logic
  programs to sets of AICs over the empty database given \S~7 of~\cite{Caroprese2011}.
  Given a logic program $\mathcal P$, deciding whether
  $\langle\emptyset,\mathit{aic}(\mathcal P)\rangle$ has a grounded repair is equivalent
  to deciding whether $\mathcal P$ has a grounded model, which is $\Sigma^P_2$-complete by
  Theorem~5.7 of~\cite{Bogaerts2015}.
\end{proof}

\begin{proof}[Lemma~\ref{lem:indep-com}]
  Assume that $\eta_1\perp\eta_2$.
  Then, for every $\U$, $\U(\DB)$ and $\T_1(\U)(\DB)$ agree on the bodies of all rules in
  $\eta_2$, so $\T_2(\U)=\U\uplus A$ and $\T_2(\T_1(\U))=\T_1(\U)\uplus A$ for some set
  $A$.
  Likewise, $\T_1(\U)=\U\uplus B$ and $\T_1(\T_2(\U))=\T_2(\U)\uplus B$, and furthermore
  $A$ and $B$ are disjoint.
  Therefore $\T_1(\T_2(\U))=\U\uplus(A\cup B)=\T_2(\T_1(\U))$.
  Furthermore, also $\T(\U)=\U\uplus(A\cup B)$.
\end{proof}

\begin{proof}[Lemma~\ref{lem:grounded-par}]
  Let $\V\subseteq\U$ and $\V_i=\V\cap\U_i\subseteq\U_i$.
  We can write $\T(\V)\cap\U$ as
  \begin{align*}
    &&\left(\T_1(\T_2(\V_1\cup\V_2))\cap\U_2\right)&\cup\left(\T_2(\T_1(\V_1\cup\V_2))\cap\U_1\right)\\
    &=&\left(\T_2(\V_1\cup\V_2)\cap\U_2\right)&\cup\left(\T_1(\V_1\cup\V_2)\cap\U_1\right)\\
    &=&\left(\T_2(\V_2)\cap\U_2\right)&\cup\left(\T_1(\V_1)\cap\U_1\right)
  \end{align*}
  where the first equality is justified by the fact that $\T_i$ can not change its input by
  elements of $\U_{\ii}$, the second by the fact that $\T_i$'s output is not affected by
  changes to elements of $\U_{\ii}$.
  We conclude that $\T(\V)\cap\U\subseteq\V$ iff $\T_i(\V_i)\cap\U_i\subseteq\V_i$ for each $i$.

  Assume $\U$ is a grounded repair for $\Ieta$, i.e.~that $\U$ is a strictly grounded
  fixpoint of $\T$.
  Since $\U$ is founded (Lemma~\ref{lem:founded}), we know by Lemma~\ref{lem:foiks}(ii)
  that $\U=\U_1\cup\U_2$ and $\U_1$ and $\U_2$ are both (founded) repairs for $\Ieta[1]$
  and $\Ieta[2]$.
  Suppose that $\V_1\subsetneq\U_1$ and that $\T_1(\V_1)\cap\U_1\subseteq\V$.
  Since $\T_2(\U_2)=\emptyset$, we conclude that $\T(\V_1\cup\U_2)\cap\U\subseteq\V$,
  which contradicts $\U$ being strictly grounded for $\T$.
  A similar argument shows that $\U_2$ is a strictly grounded fixpoint of $\T_2$.

  For the converse implication, assume that each $\U_i$ is a strictly grounded fixpoint of
  $\T_i$, and let $\V\subsetneq\U$ be such that $\T(\V)\cap\U\subseteq\V$.
  Then $\V_i\subsetneq\U_i$ for at least one of $i=1$ or $i=2$, and for that value of $i$
  it is also the case that $\T_i(\V_i)\cap\U_i\subseteq\V_i$, contradicting the fact that
  $\U_i$ is strictly grounded for $\T_i$.
\end{proof}

We omit the proof of Lemma~\ref{lem:lattice-indep}, which reduces to mechanical algebraic
manipulation of lattice identities.

\begin{proof}[Lemma~\ref{lem:grounded-strat}]
  Write $\T_1$ for $\T^\DB_{\eta_1}$ as before, but let $\T_2$ now denote
  $\T^{\U_1(\DB)}_{\eta_2}$.

  Assume that $\U$ is a grounded repair for $\Ieta$.
  As before, the equality $\U=\U_1\cup\U_2$ is a consequence of Lemmas~\ref{lem:founded}
  and~\ref{lem:foiks}(iii), and both $\U_1$ and $\U_2$ are fixpoints of $\T_1$ and
  $\T_2$.

  We show that $\U_1$ is strictly grounded for $\T_1$.
  Suppose that $\V\subsetneq\U_1$.
  Then $\T(\V\cup\U_2)\cap\U\not\subseteq\V\cup\U_2$, whence necessarily
  $\T(\V\cup\U_2)\cap\U_1\not\subseteq\V$.
  As above, $\T(\V\cup\U_2)\cap\U_1=\T_1(\V\cup\U_2)\cap\U_1=\T_1(\V)\cap\U_1$, as actions
  in $\U_1$ can only arise from rules in $\eta_1$, and the applicability of these does not
  depend on actions in $\U_2$.
  Therefore $\T_1(\V)\cap\U_1\not\subseteq\V$, and thus $\U_1$ is strictly grounded for
  $\T_1$.

  The argument for $\U_2$ is similar.
  If $\V\subsetneq\U_2$, then $\T(\U_1\cup\V)\cap\U\not\subseteq\U_1\cup\V$, whence
  now $\T(\U_1\cup\V)\cap\U_2\not\subseteq\V$.
  We now observe that $\T(\U_1\cup\V)=\T_2(\V)$, as $\U_1$ and $\V$ are necessarily
  disjoint and thus $(\U_1\cup\V)(\DB)=\V(\U_1(\DB))$.
  Therefore $\T_2(\V)\cap\U_2\not\subseteq\V$, and hence $\U_2$ is strictly grounded for
  $\T_2$.

  Now suppose that $\U_1$ and $\U_2$ are strictly grounded fixpoints of, respectively,
  $\T_1$ and $\T_2$.
  Again by Lemma~\ref{lem:foiks}(iii) we know that $\U$ is a fixpoint of $T$, so we only
  need to show that it is strictly grounded.
  Let $\V\subseteq\U$; there are two cases to consider.

  If $\V\cap\U_1\subsetneq\U_1$, then necessarily
  $T_1(\V\cap\U_1)\cap\U_1\not\subseteq\V\cap\U_1$, and since $\V\setminus\U_1$ does not
  change applicability of rules in $\eta_1$ it also follows that
  $\T_1(\V)\cap\U_1\not\subseteq\V$.
  Since the rules in $\eta_2$ cannot cancel applicability of rules in $\eta_1$, it also
  follows that $\T(\V)\cap\U\not\subseteq\V$.

  If $\V\cap\U_1=\U_1$, then necessarily $\V\cap\U_2\subsetneq\U_2$.
  In this case, we know that $\T_2(\V\cap\U_2)\cap\U_2\not\subseteq\V\cap\U_2$.
  Furthermore, the update actions in $\V\setminus\U_2=\U_1$ do not change $\U_1(\DB)$,
  hence $\T_2(\V\cap\U_2)=\T_2(\V)=\T(\V)$, where the last equality is justified from the
  fact that no rule in $\eta_1$ is applicable to $\V(\DB)$, as $\U_1$ is a repair for
  $\Ieta[1]$.
  Therefore we again conclude that $\T(\V)\cap\U\not\subseteq\V$.

  Thus for any $V\subsetneq\U$ it is the case that $\T(\V)\cap\U\not\subseteq\V$, which
  shows that $\U$ is strictly grounded for $\T$.
\end{proof}

\begin{proof}[Lemma~\ref{lem:nondet-subset}]
  \begin{align*}
    \T_\neta(\U)
    &=\U\uplus\{\head(r)\mid\U(\DB)\models\body(r),r\in\neta\} \\
    &=\U\uplus\left\{\bigcup\head(r)\mid\U(\DB)\models\body(r),r\in\eta\right\} \\
    &\subseteq\bigcup\T_\eta(\U)\qedhere
  \end{align*}
\end{proof}

\begin{proof}[Lemma~\ref{lem:nondet-grounded}]
  It suffices to show that $\left(\bigcup\T_\eta(\V)\right)\cap\U\subseteq\V$ iff
  $\T_\neta(\V)\cap\U\subseteq\V$.
  The direct implication is a direct consequence of Lemma~\ref{lem:nondet-subset}.
  The converse implication follows from the fact that any element in
  $\left(\bigcup\T_\eta(\V)\right)\setminus\T_\neta(\V)$ must anyway be in $\V$ -- as the
  example above illustrates, these are actions that are in $\V$ and are cancelled by an
  action in the head of a rule in $\eta$.
\end{proof}

\end{document}